\documentclass[]{spie}  %>>> use for US letter paper
\usepackage{amsmath,amsfonts,amssymb}
\usepackage{graphicx}
\usepackage[colorlinks=true, allcolors=blue]{hyperref}
\usepackage{setspace}
\usepackage{xcolor}
\usepackage{titlesec}
\usepackage{graphicx}
\usepackage{float}
\usepackage{subcaption}
\usepackage{booktabs}
\usepackage{enumitem}
\usepackage{comment}
\usepackage{amsmath}
\usepackage{multirow}
\usepackage{booktabs}

\title{Ultra-precise Multi-fiber Optical Connectors for Astronomy}

\author[a]{Malak Galal}
\author[a]{Maxime Rombach}
\author[a]{Jean-Paul Kneib}
\affil[a]{Institute of Physics, Laboratory of Astrophysics, Ecole Polytechnique Federale de Lausanne (EPFL), Observatoire de Sauverny, CH-1290 Versoix, Switzerland}

\authorinfo{Further author information: (Send correspondence to M. G.)\\M. G.: E-mail: malak.galal@epfl.ch, Telephone: +41 21 693 92 91\\  M. R.: E-mail: maxime.rombach@epfl.ch, Telephone: +41 21 693 28 76}

\begin{document} 
\maketitle

\begin{abstract}
The increasing sensitivity of modern astronomical instruments requires optical fiber connections with high cross-mating stability and insertion loss as low as 1$\%$ (0.05 dB). Conventional connectors, though suitable for telecommunications, introduce excess attenuation, Fresnel reflections, and alignment instabilities that degrade throughput and calibration accuracy in astronomy. This work presents the design and characterization of ultra-low-loss optical multi-fiber connectors developed for astronomical use. Fabricated using femtosecond-laser 3-D printing, they achieve sub-micron ferrule tolerances. Preliminary metrology shows excellent hole alignment and roundness, and initial throughput tests for three simultaneously connected fibers show losses as low as 0.95$\%$ (0.04 dB). Further throughput and FRD characterization to be implemented to assess efficiency, stability, and repeatability under observatory conditions.
\end{abstract}


\section{INTRODUCTION}
\label{sec:intro}  % \label{} allows reference to this section

\noindent Highly-performing optical multi-fiber connectors with exceptional reliability and minimal focal ratio degradation (FRD) have become indispensable in the era of large telescopes dealing with tens of thousands of optical fibers. Science productivity could significantly be improved if focal plane fibers and fiber bundles could have the flexibility to be reconnected as plug-and-play tools without degrading significantly the instrument’s performance. Developing reliable and repeatable fiber connectors will enable the exchange of individual fibers, fiber bundles, or various combinations between the focal plane and the highly-sensitive spectrographs which are stationary and need to be securely stored in temperature-controlled chambers. 
Currently, the most transmission efficient way to connect optical fibers to the spectrographs in astronomical instrumentation is fiber splicing—but it comes with labor cost. Next-generation large-multiplexing fiber-fed instruments require, however, the flexibility to swap fibers between focal plane arrays and different instruments, such as spectrographs operating at different wavelength ranges and resolutions, or to disconnect and reconnect fiber cables for instrument maintenance and upgrades. 
Due to the absence of specialized low-loss connectors designed specifically for astronomical applications, fiber-fed astronomical instruments have typically adapted existing telecommunications connectors instead of developing dedicated solutions. However, commercially available fiber connectors often lack the robustness needed for astronomy and exhibit higher losses than desired. Among the available options, the Multi-fiber Termination Push-on (MTP) connectors by US Conec have been widely studied for their potential use in astronomical applications. While these connectors are optimized for telecommunications, they have not been specifically tailored to meet the demanding requirements of astronomical instrumentation and generally exhibit insertion losses which are not consistent and may reach more than 10$\%$ of loss. 
The goal of this work is to develop fiber connectors that operate across the 370–2400 nm wavelength range while maintaining losses comparable to fiber splicing (0.05 dB corresponding to ~1$\%$ light loss). This means that new developments need to be carried out in order to attain this performance. To that effect, we aim to design two sorts of highly-precise multi-fiber optical connectivity components namely, the symmetric multi-fiber connector and the rotary optical switch. Both types of connectivity components will be especially important for the next-generation multi-object spectroscopy astronomical surveys, like the Chinese MUST \cite{zhao2024multiplexed}, the American Spec-S5 \cite{besuner2025spectroscopic}, and the European WST \cite{mainieri2024wst}, that will utilize several tens of thousands of optical fibers to collect skylight and transmit it to the spectrographs. Our novel approach consists in utilizing ultra-precise manufacturing of our meticulous mechanical designs and 3-D printing the parts using silica glass substrates. Commercially-available connectors are made of polymer, and that explains the reason why such connectors do not reach the high precision needed for astronomy. 
Ultra-precision manufacturing in silica glass is the breakthrough that optical connectivity components currently need. Together with our rigorous mechanical designs, we will be able to assemble astronomy-grade connectivity components that are ultra-precise in alignment, and in turn will exhibit extremely low-losses reaching the required 1$\%$ level. Additionally, our careful designs will minimize FRD losses, and with our strong expertise in optical testing, we will be able to identify and address any problems/risks and will mitigate them promptly.

\section{MULTI-OBJECT SPECTROSCOPY AND ITS RELIANCE ON FIBER OPTICS}

\begin{figure}[!b]
    \centering
    \includegraphics[width=0.7\linewidth]{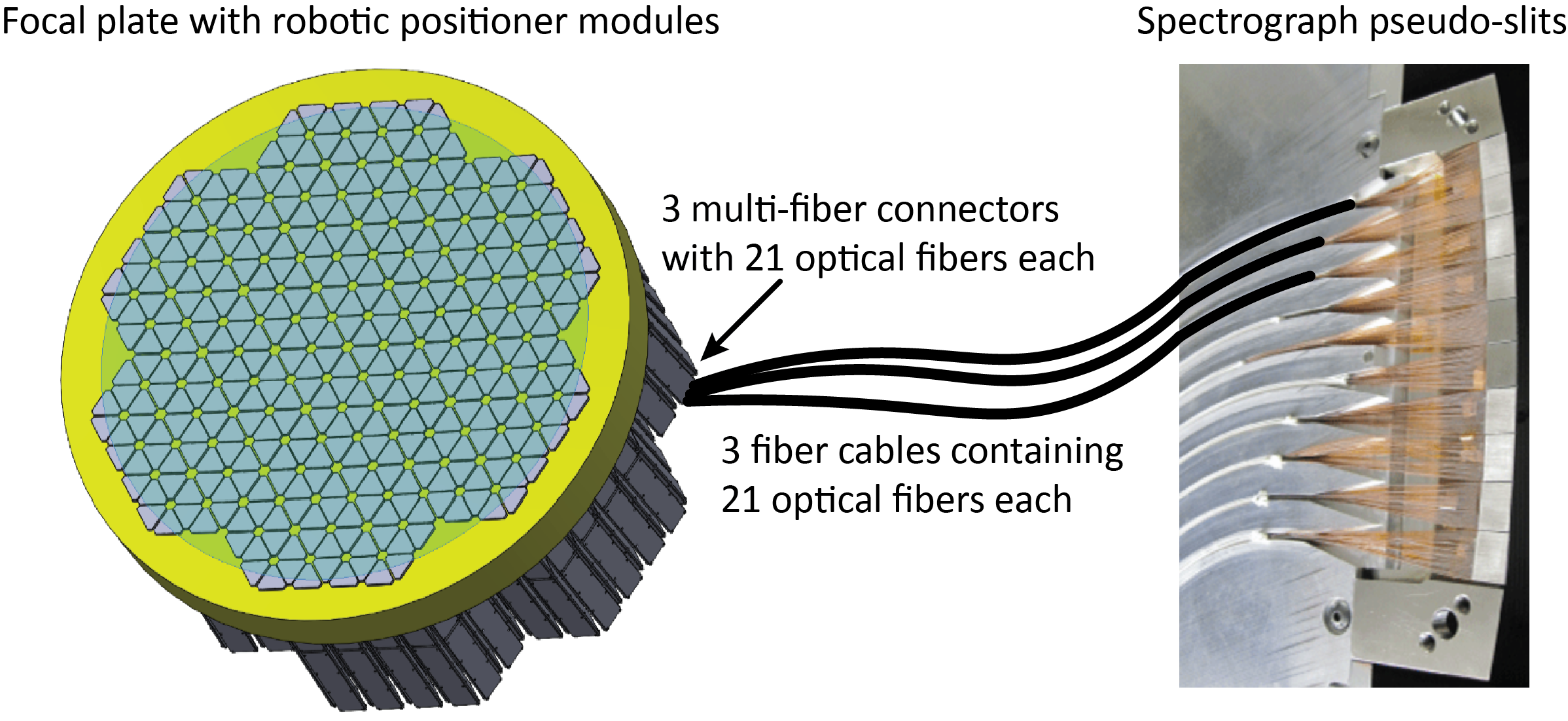}
    \caption{Schematic showing the focal plate with modules each having 63 robotic positioners and each positioner holding one optical fiber to be routed and be placed onto V-groves of the spectrograph pseudo-slits (Picture of pseudo-slits adapted from the APOGEE spectrograph \cite{wilson2019apache}).}
    \label{fig:platewithmodules}
\end{figure}

\noindent Multi-Object Spectroscopy (MOS) represents a particular kind of astronomical surveys relying on employing optical fibers to observe a large number of sky objects simultaneously \cite{smee2013multi}. The optical fibers are placed within the focal plate of the telescope to collect skylight and transmit it to the spectrograph where the data is analyzed. The spectrograph is quite a sensitive and fragile piece of equipment that needs to be kept safely within temperature-controlled chambers and accordingly cannot be directly placed by the focal plate. To that effect, a large number of several meters of optical fibers will be carefully routed from the focal plate to reach the spectrograph (as shown in Figure \ref{fig:platewithmodules}) with as minimal optical loss as possible.

One of the current major running astronomical surveys is the Dark Energy Spectroscopic Instrument (DESI) \cite{dey2019overview}. It has in its focal plate 5k robotic positioners each holding an optical fiber of about 50 m running to the spectrograph where they are fixed onto fiber slits. For this survey, the fibers have been spliced \cite{poppett2024overview}, as, thus far, there are no optical connectors or switches available that meet the requirements of the project. The splicing loss of optical fibers is about 0.05 dB (1$\%$ of light loss), and an astronomical survey would require ideally connectors exhibiting the very same order of loss or even less \cite{poppett2024overview}.
Although fiber splicing achieves very low losses, it remains a time-consuming process that demands specialized equipment, consumables, and skilled personnel. In addition, the need for testing and potential re-splicing during maintenance further increases overall costs. By contrast, the development of advanced multi-fiber connector technology offers a promising alternative. Since each connector can accommodate multiple fibers, the overall cost of implementing such a system is expected to be significantly lower than that of traditional splicing, making connectors a more economical and practical solution in the long term. For the current 5,000 optical fibers, it is still manageable to use splicing, but with 20,000 and more fibers, it will be quite difficult to rely on splicing, and the concept of modularity and ease of maintenance will also push for having fiber connectors.

Multi-fiber optical connectors have been extensively utilized in high-density fiber optic networks (e.g. data centers) to facilitate the linking of servers to optical switches and other networking devices. Multi-fiber Push-On (MPO) and Multi-fiber Termination Push-on (MTP) connectors are at the heart of modern optical fiber communication systems supporting the infrastructure for high-speed data networks. Even though the necessity for high-capacity and highly efficient data transmission is becoming more important than ever due to the rapid development of technology, telecom-grade multi-fiber connectors and optical switches are no way near to the required performances set by astronomical surveys \cite{poppett2024overview, sathi2024optimizing}.
Currently the best available optical connectors, such as the MTP connectors by the company US Conec, show inconsistency in their loss characteristics with losses ranging from 10$\%$ to 30$\%$ \cite{sathi2024optimizing, farr2022fiber}, which is quite far from what is needed for astronomical surveys. These significantly high losses result from the severe misalignments to the opposite fiber holes, alignment pins, etc. mainly caused by imprecise mass-manufacturing \cite{suematsu2003super}. 

\section{FERRULE MATERIAL AND DESIGN ARCHITECTURE} \label{sec:design}
\begin{figure}[b]
   \centering
   \includegraphics[width=0.75\textwidth]{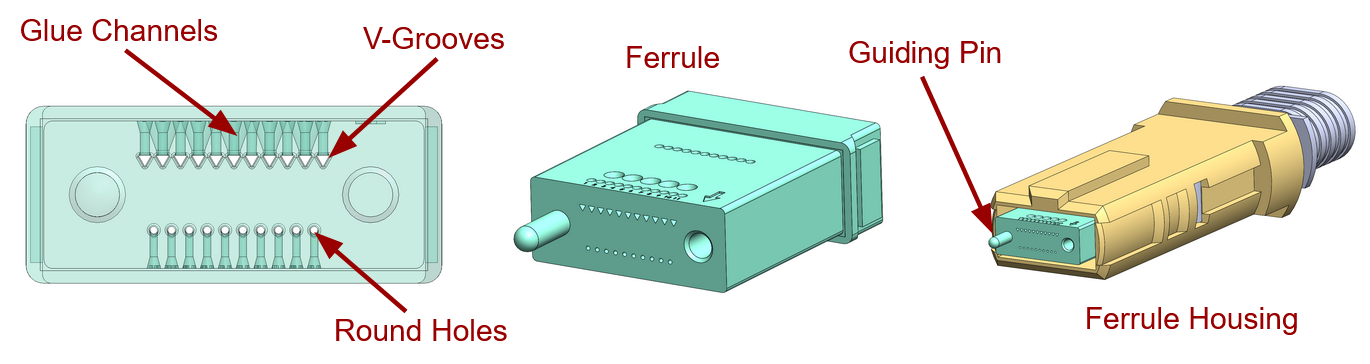}
   \caption{CAD Model of the symmetric-mating ferrule showing the unified mechanical architecture incorporating the guiding pin, guiding hole, internal fiber V-grooves, and engineered glue channels.}
   \label{fig:ferrule_cad}
\end{figure}

\noindent Our novel approach consists in utilizing ultra-precise manufacturing of our meticulous mechanical designs and 3-D printing the parts using silica glass substrates. The silica symmetric-mating multi-fiber connector is engineered to optimize fiber alignment by pairing exactly one guiding pin and one guiding hole across a symmetric mating interface. By precisely micro-machining these components together in silica glass, matching fiber hole coordinates are guaranteed to eliminate structural core-to-core lateral offsets. This architecture integrates specific mechanical elements as highlighted in the CAD model in Figure \ref{fig:ferrule_cad}). The dimensions of the ferrules are similar to conventional MTP ones, so that they can easily fit in the conventional housing.

\begin{itemize}
    \item \textbf{Fiber Round Holes and V-grooves:} Specialized geometries to hold and guide raw unclad fibers securely.
    \item \textbf{Guiding Pin and Guiding Hole:} Self-contained matching components ensuring high core-to-core co-axial alignment stability.
    \item \textbf{Glue Channels:} Micro-engineered cavities intended for adhesive distribution without distorting fiber alignment or spilling onto the optical mating face.
\end{itemize}

\begin{figure}[t]
   \centering
   % --- SUBFIGURE 1: UNCROPPED PHYSICAL PROTOTYPE ---
   \begin{subfigure}[b]{0.45\textwidth}
       \centering
       \includegraphics[width=\textwidth]{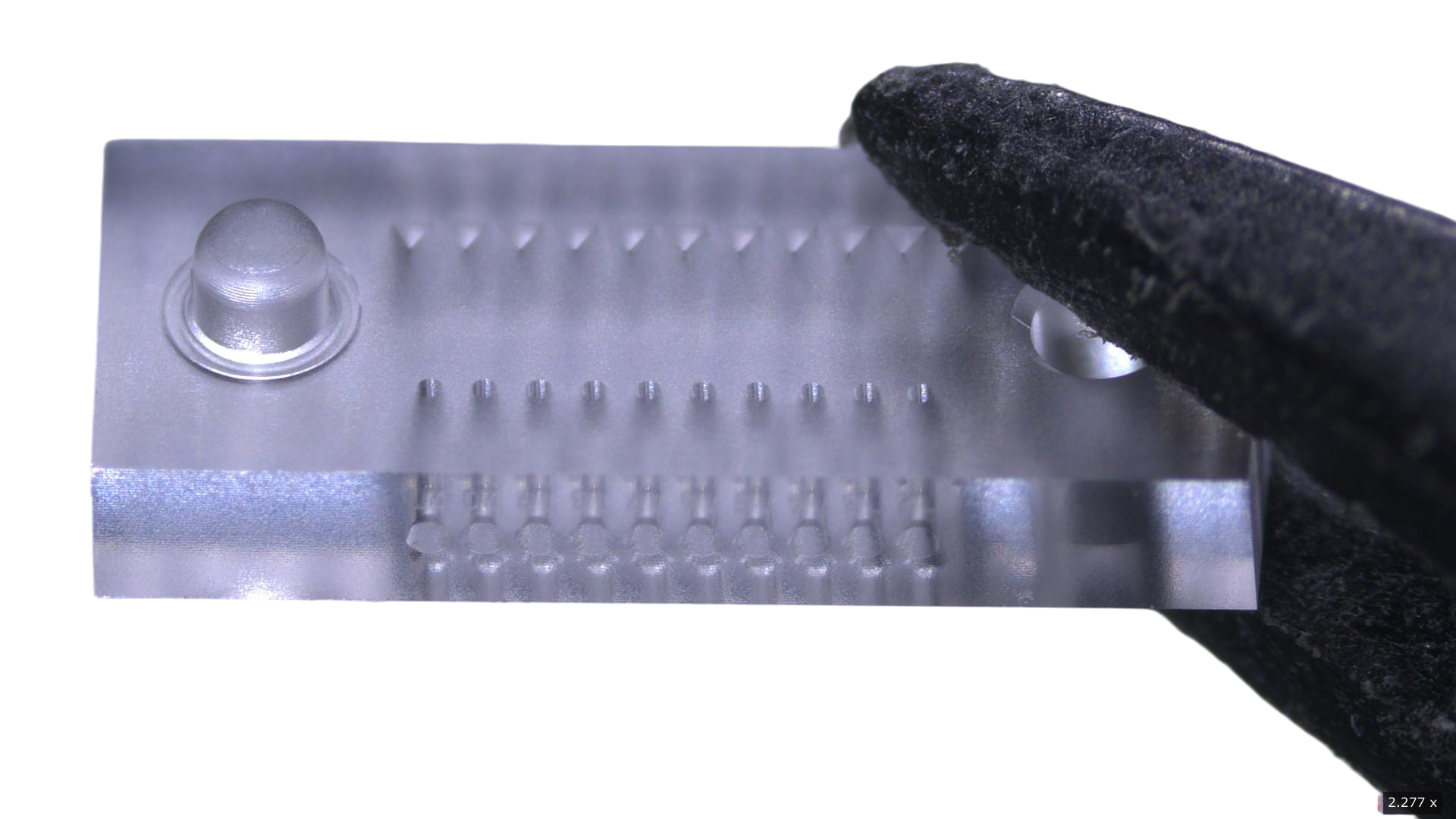}
       \caption{}
       \label{fig:ferrule_tweezer}
   \end{subfigure}\hspace{1.0cm}% <--- Explicit gap size; NO empty lines directly below this
   % --- SUBFIGURE 2: CROPPED MANUAL ASSEMBLY ---
   \begin{subfigure}[b]{0.2\textwidth} % Scaled up to 0.26 to closely match the height of the first panel
       \centering
       \includegraphics[width=\textwidth, trim={0cm 5.5cm 0cm 10.5cm}, clip]{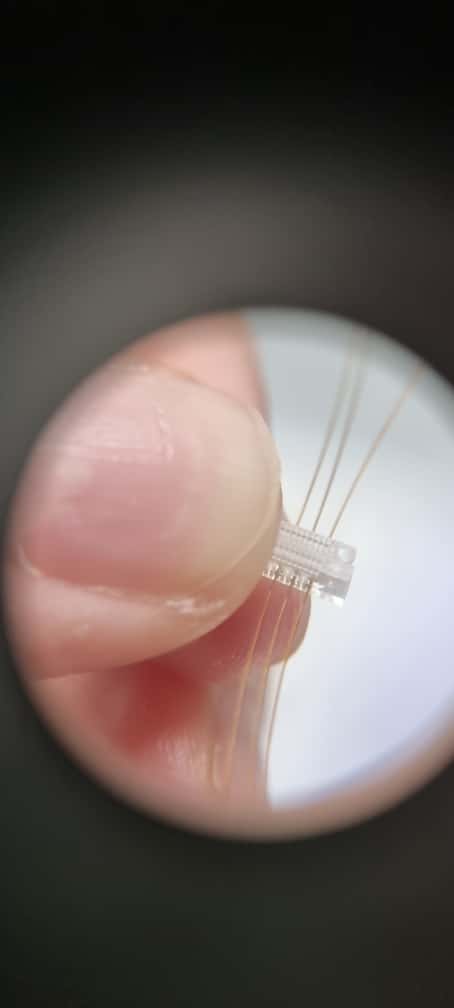}
       \caption{}
       \label{fig:ferrule_cropped_vertical}
   \end{subfigure}
   \caption{(a) A real image captured by FEMTOprint of the silica ferrule; (b) Two mated ferrules with three optical fibers each.}
   \label{fig:real_ferrule}
\end{figure}

The physical prototypes were successfully fabricated by FEMTOprint SA in Switzerland using femtosecond-laser 3-D printing as shown in Figure \ref{fig:real_ferrule}. This first round of prototyping included both the concept of round holes and V-grooves for the sake of testing. Figure \ref{fig:ferrule_tweezer} shows a real image of the silica ferrule captured by the manufacturer, and Figure \ref{fig:ferrule_cropped_vertical} shows two silica ferrules mated with three optical fibers on in each ferrule.

\begin{figure}[b]
   \centering
   \begin{subfigure}[b]{0.4\textwidth}
       \centering
       \includegraphics[width=\textwidth]{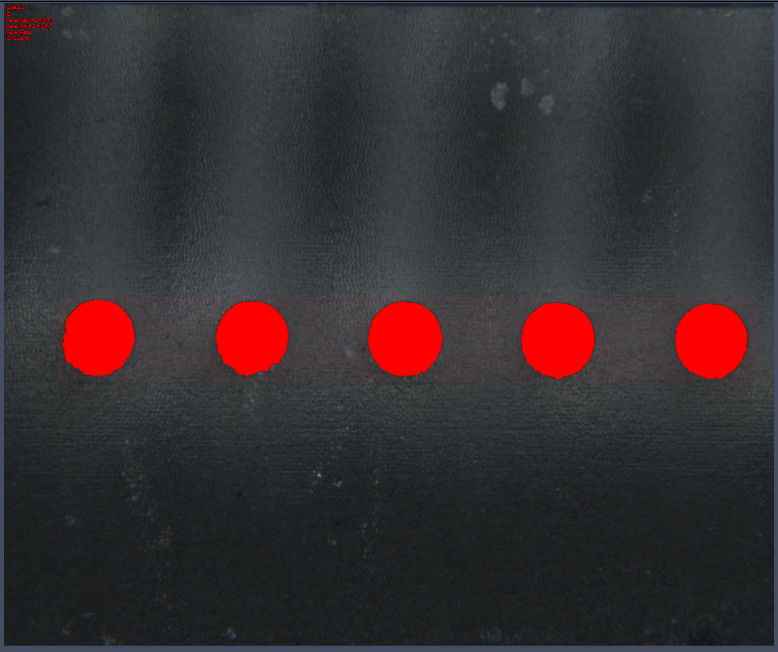}
       \caption{FEMTOprint silica ferrule}
       \label{fig:femtoprint_threshold}
   \end{subfigure}
   \hspace{1.5cm}
   \begin{subfigure}[b]{0.4\textwidth}
       \centering
       \includegraphics[width=\textwidth]{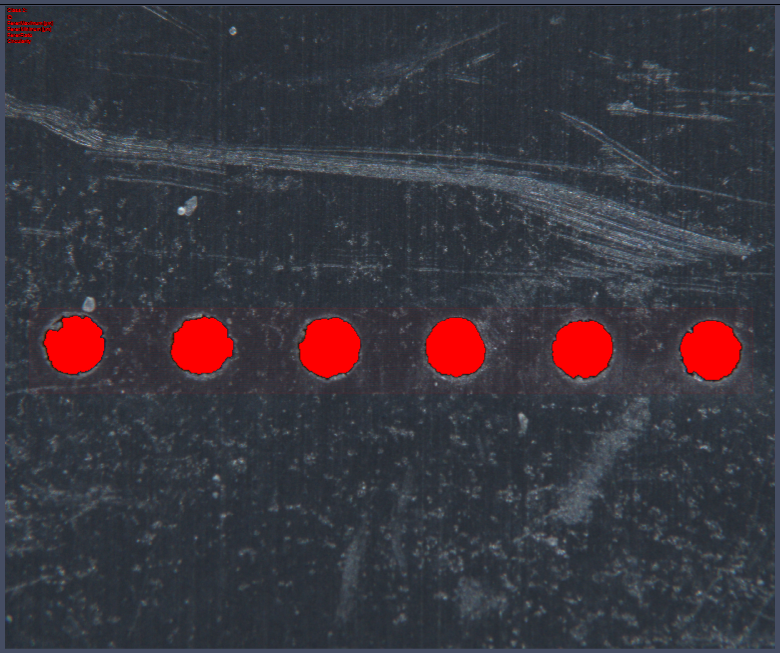}
       \caption{Conventional MTP polymer ferrule}
       \label{fig:mtp_threshold}
   \end{subfigure}
   \caption{High-magnification (112x) micro-hole inspections utilizing binary thresholding, comparing the structural roundness and alignment of the custom fabricated silica glass ferrule (a) against a standard commercial polymer connector (b).}
   \label{fig:threshold_comparison}
\end{figure}

Preliminary verification under a Zeiss optical microscope verified pristine micro-hole tolerances and high-magnification threshold inspections (112x binary thresholding) confirmed strict concentric hole circularity and dimensional alignment. A comparison between the circularity and smoothness of the holes can be seen in Figure \ref{fig:threshold_comparison}. The figure clearly shows that the MTP ferrule suffers from polymer residual inside the holes, whereas the FEMTOprint ferrule looks more round and smooth.

To quantify the geometric advantages of the laser 3D-printed silica substrate over conventional injection-molded alternatives, an automated center-of-mass metrology analysis was performed on the thresholded microscope profiles shown in Figure \ref{fig:threshold_comparison}. The custom silica glass ferrule demonstrated superior spatial uniformity as shown in Table \ref{tab:calibrated_pitch_metrology}, yielding a channel-to-channel pitch standard deviation of just $\pm$0.32 pixels ($\pm$0.52 $\mu$m). By comparison, the commercial polymer MTP connector exhibited a spacing standard deviation of $\pm$1.00 pixels ($\pm$1.97 $\mu$m), which is over three times higher than the silica substrate. This pronounced spatial variance, combined with noted edge irregularities in the polymer channels, confirms that conventional mass-manufacturing tolerances are insufficient for the sub-micron alignment precision required by next-generation multi-object astronomical spectrographs.

% \begin{comment}
% \begin{figure}[b]
%    \centering
%    \includegraphics[width=0.75\textwidth]{Figures/Ferrule with Tweezer.png}
%    \caption{Silica glass ferrule precisely manufactured by FEM}
%    \label{fig:ferrule_cad}
% \end{figure}
% \end{comment}

\begin{table}[htbp]
\caption{Calibrated step-by-step channel pitch spacing and manufacturing uniformity comparison for the custom 3-D-printed silica glass ferrule and conventional polymer MTP connector.}
\label{tab:calibrated_pitch_metrology}
\begin{center}       
\begin{tabular}{ccc}
\toprule
\textbf{Measurement Parameter} & \textbf{Silica Glass Ferrule ($\mu$m)} & \textbf{Polymer MTP Connector ($\mu$m)} \\ 
\midrule
Space between Hole 1 and 2  & 250.68 & 248.49 \\
Space between Hole 2 and 3  & 249.93 & 251.93 \\
Space between Hole 3 and 4  & 249.23 & 247.13 \\
Space between Hole 4 and 5  & 250.00 & 250.18 \\
Space between Hole 5 and 6  & ---    & 252.28 \\ 
\midrule
\textbf{Standard Deviation ($\sigma$)} & \textbf{$\pm$0.52 $\mu$m} & \textbf{$\pm$1.97 $\mu$m} \\
\bottomrule
\end{tabular}
\end{center}
\end{table}
\section{EXPERIMENTAL THROUGHPUT PERFORMANCE} \label{sec:results}
\noindent Empirical validation of the multi-fiber connection was conducted by inserting optical fibers into the 3D-printed silica ferrule channels and establishing mating pairings. Optical throughput baseline references were established by measuring absolute power on a precision fusion splicer equipped with micro-step manual alignment capabilities. The connection efficiency was subsequently quantified by recording absolute transmitted power across sequential mating cycles through the connected glass ferrule interfaces. The measured throughput and corresponding calculated optical losses are compiled in Table~\ref{tab:throughput}.

\begin{comment}
\begin{table}[h]
\caption{Experimental throughput measurements and current insertion loss values recorded across sequential mating cycles for three optical fibers.} 
\label{tab:throughput}
\begin{center}       
\begin{tabular}{ccccc}
\toprule
\textbf{\begin{tabular}[c]{@{}c@{}}Fiber\\ Number\end{tabular}} & \textbf{\begin{tabular}[c]{@{}c@{}}Mating\\ Trial\end{tabular}} & \textbf{\begin{tabular}[c]{@{}c@{}}Splicer Reference\\ Power (W)\end{tabular}} & \textbf{\begin{tabular}[c]{@{}c@{}}Connected Ferrule\\ Power ($\mu$W)\end{tabular}} & \textbf{\begin{tabular}[c]{@{}c@{}}Current Measured\\ Optical Loss\end{tabular}} \\ \midrule
\multirow{2}{*}{\textbf{1}} & Mate 1 & \multirow{2}{*}{2.11} & 2.09 & 0.95\% (0.04 dB loss) \\
                            & Mate 2 &                       & 2.06 & 2.37\% (0.11 dB loss) \\ \midrule
\multirow{2}{*}{\textbf{2}} & Mate 1 & \multirow{2}{*}{3.37} & 3.31 & 1.78\% (0.08 dB loss) \\
                            & Mate 2 &                       & 3.28 & 2.67\% (0.12 dB loss) \\ \midrule
\multirow{2}{*}{\textbf{3}} & Mate 1 & \multirow{2}{*}{8.61} & 8.08 & 6.15\% (0.28 dB loss) \\
                            & Mate 2 &                       & 7.95 & 7.66\% (0.32 dB loss) \\ 
\bottomrule
\end{tabular}
\end{center}
\end{table}
\end{comment}

% Ensure you have \usepackage{booktabs} and \usepackage{multirow} in your main.tex preamble

\begin{table}[h]
\caption{Experimental optical throughput efficiency and corresponding insertion loss values recorded across sequential mating cycles for three simultaneously connected fibers using the FEMTOprint silica ferrules.} 
\label{tab:throughput}
\begin{center}       
\begin{tabular}{cccc}
\toprule
\textbf{Fiber Number} & \textbf{Mating Trial} & \textbf{Throughput Efficiency (\%)} & \textbf{Measured Insertion Loss (dB)} \\ 
\midrule
\multirow{2}{*}{\textbf{1}} & Mate 1 & 99.05\% & 0.04 dB \\
                            & Mate 2 & 97.63\% & 0.11 dB \\ 
\midrule
\multirow{2}{*}{\textbf{2}} & Mate 1 & 98.22\% & 0.08 dB \\
                            & Mate 2 & 97.33\% & 0.12 dB \\ 
\midrule
\multirow{2}{*}{\textbf{3}} & Mate 1 & 93.85\% & 0.28 dB \\
                            & Mate 2 & 92.34\% & 0.32 dB \\ 
\bottomrule
\end{tabular}
\end{center}
\end{table}

The initial test data proves highly encouraging. For both Fiber 1 and Fiber 2, insertion losses successfully approach the near-zero performance threshold targets. Specifically, Fiber 1 during its first mating cycle achieved an ultra-low loss of 0.95\% (0.04 dB), which performs symmetrically on par with an ideal permanent fusion splice ($\sim$0.05 dB). 

\section{DISCUSSION AND CONCLUSION} \label{sec:discussion_conclusion}
\noindent The structural and empirical results presented in this work demonstrate the disruptive potential of micro-machined silica glass ferrules over conventional injection-molded polymer alternatives for high-density astronomical instrumentation. Next-generation multi-object spectroscopic surveys require sub-micron cross-mating alignment to keep insertion attenuation near the 1\% (0.05~dB) splice-equivalent benchmark. Automated threshold metrology confirms that conventional polymer MTP blocks fall short of these strict tolerances, exhibiting an erratic channel-to-channel pitch standard deviation of $\pm$1.00 pixel ($\pm$1.97~$\mu$m), which directly correlates with the inconsistent 10\% to 30\% throughput losses documented in the literature. In stark contrast, the laser 3-D-printed silica ferrules achieve a sub-micron pitch uniformity deviation of just $\pm$0.32 pixels ($\pm$0.52~$\mu$m), structurally locking the fiber core coordinates across the mating interface.

Beyond superior alignment stability, the femtosecond-laser micro-machining process yields profound advantages regarding internal geometric roundness. High-magnification microscopy reveals that polymer micro-injection molding leaves behind micro-warping, jagged internal walls, and residual material burrs. In fiber-fed astronomical systems, these focal perturbations scatter propagating light paths, degrading the output numerical aperture and causing catastrophic FRD. Conversely, the pristine, smooth internal surfaces of the silica glass micro-holes support the optical fiber uniformly without localized stress points. This clean mechanical interface is expected to yield exceptional FRD preservation alongside near-zero throughput loss.

This structural integrity is empirically validated by our initial multi-fiber throughput characterizations. Fiber 1 and Fiber 2 achieved outstanding initial peak transmission efficiencies of 99.05\% (0.04~dB loss) and 98.22\% (0.08~dB loss), respectively. 

In conclusion, ultra-precise manufacturing in silica glass delivers the structural breakthrough required by next-generation multi-fiber astronomical instruments. To mature this technology for observatory integration, more comprehensive throughput testing using automated robotic alignment rigs will verify long-term mating repeatability, mechanical stability, and environmental thermal resilience. Furthermore, future work will focus on executing a a thorough focal ratio degradation characterization to measure light cone conservation under varying stress states. Additionally, the development of a specialized polishing jig required to achieve optical-grade, coplanar glass surface finishes across all fiber channels simultaneously remains under active investigation.

\acknowledgments % equivalent to \section*{ACKNOWLEDGMENTS}       
 
The authors would like to thank the EPFL Equipment Fund for supporting this work.

% References
\bibliography{report} % bibliography data in report.bib
\bibliographystyle{spiebib} % makes bibtex use spiebib.bst

\end{document}